# Enhanced thermoelectric performance of carbon nanotubes at elevated temperature

P. H. Jiang, H. J. Liu[*], D. D. Fan, L. Cheng, J. Wei, J. Zhang, J. H. Liang, J. Shi

*Key Laboratory of Artificial Micro- and Nano-structures of Ministry of Education and School of Physics and Technology, Wuhan University, Wuhan 430072, China*

The electronic and transport properties of (10, 0) single-walled carbon nanotube are studied by performing the first-principles calculations and semi-classical Boltzmann theory. It is found that the (10, 0) tube exhibits considerably large Seebeck coefficient and electrical conductivity which is highly desirable for good thermoelectric materials. Together with the lattice thermal conductivity predicted by non-equilibrium molecular dynamics simulations, the room temperature *ZT* value of (10, 0) tube is estimated to be 0.15 for *p*-type carriers. Moreover, the *ZT* value exhibits strong temperature dependence and can be reached to 0.77 at 1000 K. Such *ZT* value can be further enhanced to as high as 1.7 by isotope substitution and chemisorptions of hydrogen on the tube.

## 1. Introduction

Thermoelectric materials can directly convert heat into electricity and vice verse, which provide a new way to ease the energy crisis and can be also used as refrigeration devices. The efficiency of a thermoelectric material is characterized by the dimensionless figure of merit

$$ZT = \frac{S^2 \sigma T}{\kappa_e + \kappa_l}, \qquad (1)$$

where $S$ is the Seebeck coefficient, $\sigma$ is the electrical conductivity, $T$ is the absolute temperature, $\kappa_e$ and $\kappa_l$ are the electronic thermal conductivity and lattice thermal conductivity, respectively. To compete with traditional energy conversion methods, the *ZT* value of a thermoelectric material should be larger than 3.0. Nevertheless, such target value is still far from being reached, which can be attributed to the fact that the transport coefficients in Eq. (1) are coupled with each other and it is extremely difficult to greatly improve the *ZT* value [1]. In 1993, Hicks and Dresselhaus predicted that the *ZT* value of low-dimensional structures can be significantly enhanced due to the quantum confinement effect and surface phonon scattering [2, 3]. These pioneering studies have stimulated a lot of subsequent works, both theoretically and experimentally, and improvement of the thermoelectric performance has been confirmed in variety of systems, such as $Bi_2Te_3$/$Sb_2Te_3$ superlattice structure [4], PbSeTe/PbTe quantum-dot superlattice [5], and Si nanowires [6].

---
[*] Author to whom correspondence should be addressed. Electronic mail: phlhj@whu.edu.cn

As an interesting quasi-one-dimensional material, carbon nanotube (CNT) is found to exhibit outstanding electronic transport properties [7, 8]. In the community of thermoelectric research, CNTs are usually used as additive to improve the thermoelectric performance of organic composites [9, 10]. However, the thermoelectric properties of CNT itself are less studied. In fact, several earlier attempts indicated that the *ZT* values of CNTs are rather small. For example, Kunadian *et al.* [11] experimentally reported a room temperature *ZT* value of about $5.0\times10^{-5}$ for B-doped and N-doped multi-walled carbon nanotube. Zhao *et al.* [12] observed that the Ar plasma treatment could improve the thermoelectric performance of CNT bulky papers significantly, and the *ZT* value can be raised from 0.01 to 0.4. Theoretical approach using non-equilibrium Green's function (NEGF) method predicted that the *ZT* value of (10, 0) single-walled carbon nanotube (SWCNT) is 0.14 at 300 K [13]. In consideration of the outstanding electronic transport properties, the fairly small *ZT* values of CNTs should be originated from their quite large lattice thermal conductivity, which ranges from 78 W/mK to 7000 W/mK as reported by different groups [14, 15, 16, 17, 18]. As a comparison, the lattice thermal conductivity of the typical thermoelectric material $Bi_2Te_3$ is only ~ 2.0 W/mK [19]. If CNT could be used as a realistic thermoelectric material, the thermal conductivity should be effectively reduced while maintaining or less affecting its good electronic transport properties.

In this work, using a multi-scale approach combining first-principles calculations, semi-classical Boltzmann theory, and molecular dynamics (MD) simulations, we provide a systematical investigation on the thermoelectric properties of (10, 0) SWCNT. We demonstrate that at room temperature, the $ZT$ value of (10, 0) tube is indeed very small. More importantly, we show that the $ZT$ value exhibit a strong temperature dependence, which is absent in previous works. At elevated temperature of 1000 K, the $ZT$ value of (10, 0) tube can be significantly enhanced to 1.7 by isotope substitution and chemisorptions of hydrogen, which makes CNT a promising thermoelectric material containing earth-abundant and environmental-friendly elements.

## 2. Computational details

The structure optimization and energy band of SWCNT are calculated by using first-principles project augmented wave (PAW) method [20, 21, 22]. The exchange correlation energy is in the form of Perdew-Bruke-Ernzerhof (PBE) [23], and we also consider a more accurate hybrid density functional in the form of Heyd-Scuseria-Ernzerhof (HSE) [24]. We adopt a rectangular supercell where the closest distance between the tube and its periodic images is set to 18 Å to avoid tube-tube interactions. The **k** points are sampled on a uniform grid along the tube axis, and the cutoff energy is 400 eV. The system is fully relaxed and the atomic positions are determined until the magnitude of the force acting on each carbon atom becomes less than 0.01 eV/Å. Based on the energy band structure, the Seebeck coefficient $S$ and the electrical conductivity $\sigma$ are obtained by performing the semi-classical Boltzmann theory [25], where the relaxation time is estimated from the deformation potential (DP) theory [26]. The electronic thermal conductivity $\kappa_e$ is calculated according to the Wiedemann-Franz law [27].

For the phonon transport, the lattice thermal conductivity is estimated by performing the non-equilibrium molecular dynamics (NEMD) simulations, where a Tersoff potential [28] is used to describe carbon-carbon interactions and the time step is set to 0.5 fs. To make sure the system has reached a steady state, the system is first simulated in a *NVT* ensemble for several hundred picoseconds, and then switched into a *NVE* ensemble for another several hundred picoseconds. The tube is divided into 40 equal segments where the cold and hot regions are located in the 1st and 21st segments. The kinetic energy of the coldest atoms in the hot region and the hottest atoms in the cold region are exchanged every hundreds steps according to the Müller-Plathe algorithm [29]. A steady temperature gradient and heat flux are then realized, and the lattice thermal conductivity is calculated by the Fourier's law.

## 3. Results and discussion

According to its chiral indices, it is well known that (10, 0) SWCNT is a semiconductor. Figure 1 plots the energy band structure with 100 uniform **k** points along the tube axis. We see that both the valence band maximum (VBM) and the conduction band minimum (CBM) appear at the $\Gamma$ point and are doubly degenerate. The band gap is calculated to be 0.75 eV using PBE functional. As the traditional density functional theory (DFT) tends to underestimate the band gap seriously, we repeat the band structure calculation by using more accurate HSE functional. The noticeable difference between these two band structures is that the HSE calculation upshifts both the valence and conduction bands. However, the shift of the latter is much larger, which leads to an increase of the band gap from 0.75 eV to 0.93 eV. Such gap value is very close to that obtained by using a linear combination of atomic orbitals formalism [30], and larger than other standard DFT calculations with local density approximation (LDA) or generalized gradient approximation (GGA) [31, 32]. Meanwhile, we find that HSE calculation also changes the band shapes. In particular, the slopes of energy bands near VBM are larger than those calculated with PBE functional, which gives reduced effective mass and in turn affects the electronic transport properties.

Based on the energy band structure, we are able to derive the electronic transport coefficients by using the semi-classical Boltzmann theory with relaxation time approximation [25]. In this approach, the Seebeck coefficient $S$ is independent with the relaxation time $\tau$, while the electrical conductivity $\sigma$ can only be calculated with respect to $\tau$. The accurate evaluation of the relaxation time is very complicated since it depends on the detailed scattering mechanisms. It is fortune that the scheme can be simplified since the carriers will be scattered predominantly by acoustic phonon [33]. In this regard, we adopt the DP method [26] to evaluate the relaxation time which can effectively describe the electron-acoustic phonon scattering process. For one-dimensional system such as the CNT, the relaxation time can be expressed as

$$\tau = \frac{\hbar^2 C_{1D}}{(2\pi k_B T)^{1/2} |m^*|^{1/2} E_{1D}^2}, \qquad (2)$$

where $C_{1D}$ is the elastic constant, $m^*$ is the density-of-states effective mass, and $E_{1D}$ is the absolute deformation potential (ADP) constant which is obtained by choosing the average electrostatic potential as the reference energy level [34]. These quantities can be derived from first-principles calculations. The other parameters in Eq. (2) are the temperature $T$, the reduced Planck constant $\hbar$, and the Boltzmann constant $k_B$. In addition, the band degeneracy should be considered when calculating the effective mass [35]. Table I summarizes our calculated results at room temperature, where we find that the relaxation time of (10, 0) tube is more or less consistent with those derived from the mean free path of larger semiconducting SWCNTs [36]. On the other hand, it is obviously larger than those of the typical thermoelectric materials such as $Bi_2Te_3$ ($2.2\times10^{-14}$s for *p*-type system at 300 K) [37], which is very beneficial for improving the thermoelectric performance.

In Figure 2, we show the electronic transport coefficients of (10, 0) tube as a function of carrier concentration at 300 K. For one-dimensional materials such as CNT, the carrier concentration depends on how to define the cross-sectional area, which has some arbitrariness by taking CNT as a solid cylindrical [18], or a hollow cylindrical with thickness of 1.44 Å [14] or 3.40 Å [17]. In this work, we use the solid cylindrical model so that the cross-sectional area $A = \pi d^2 / 4$, where $d$ is the diameter of CNT [18]. For both *p*-type and *n*-type carriers, the absolute Seebeck coefficient $S$ of (10, 0) tube shown in Fig. 2(a) decreases with increasing carrier concentration and becomes vanished when the concentration is larger than $10^{21}$ cm$^{-3}$. Note that at a carrier concentration of $2.0\times10^{20}$ cm$^{-3}$, the Seebeck coefficient is calculated to be 195 µV/K which agrees well with the experimentally measured result for individual SWCNTs [38]. On the other hand, we see from Fig. 2(b) that the electrical conductivity $\sigma$ increases significantly when the carrier concentration is larger than $10^{20}$ cm$^{-3}$, but maintains a smaller value at the concentration range where the Seebeck coefficient is large enough. To balance the conflicting behaviors of the Seebeck coefficient and the electrical conductivity, we should therefore try to find an optimal carrier concentration so that the power factor ($S^2\sigma$) in Eq. (1) can be maximized. Indeed, we see from Fig. 2(c) that the power factor of (10, 0) tube exhibit peak values for both *p*-type and *n*-type carriers. In particular, at a concentration of $2.1\times10^{20}$ cm$^{-3}$, a *p*-type power factor of 0.46 W/mK$^2$ can be achieved which is more than one order of magnitude higher than that of bulk Bi$_2$Te$_3$ [37]. Such favorable electronic transport properties again suggest possible thermoelectric applications of CNT.

We now move to the discussions of heat transport, which contains contributions from both electrons and phonons. The electronic thermal conductivity can be derived from the electrical conductivity according to the Wiedemann-Franz law [27]

$$\kappa_e = L\sigma T, \qquad (3)$$

Here the Lorenz number $L$ for one-dimensional system is expressed as [3]

$$L = \left(\frac{k_B}{e}\right)^2 \left[\frac{5F_{3/2}}{F_{-1/2}} - \left(\frac{3F_{1/2}}{F_{-1/2}}\right)\right]^2, \qquad (4)$$

where $F_i$ is an integration of the reduced Fermi energy $\eta = E_f / k_B T$ given by

$$F_i = F_i(\eta) = \int_0^\infty \frac{x^i dx}{e^{x-\eta} + 1}. \qquad (5)$$

The calculated electronic thermal conductivity of (10, 0) tube shows similar behavior as electrical conductivity and is thus not shown here.

For the lattice thermal conductivity $\kappa_l$, we have performed the NEMD simulation where $\kappa_l$ can be derived from the Fourier's law

$$\kappa_l = \frac{J}{A \cdot \nabla T}. \tag{6}$$

Here $J$ is the heat flux, and $\nabla T$ is the temperature gradient. The cross-sectional area $A$ is defined to be the same as that of electronic transport coefficients, so that the ZT value given by Eq. (1) does not depend on the arbitrary definition of the area. It should be noted that a quantum correction for the MD temperature should be considered [39, 40] when the simulation temperature is lower than the Debye temperature, which is 2400 K for CNT [41]. Moreover, when dealing with the lattice thermal conductivity of low-dimensional system, one should pay special attention to the size effect. It is reported that $\kappa_l$ of CNT increases with the tube length $L$ in a large range of $L$ [42, 43]. To obtain a realistic $\kappa_l$ of our (10, 0) tube, we calculate $\kappa_l$ at a series of tube length $L$ that ranges from 0.04 μm to 2.1 μm which is longer than the phonon mean free path of SWCNT [44]. The results can be fitted by an exponential law

$$\kappa_l = a e^{-L/\beta} + \kappa_0, \tag{7}$$

where $a = -666$ W/mK, $\beta = 0.366$ μm and $\kappa_0 = 891$ W/mK. For any realistic CNT samples, we can extrapolate the length $L$ to infinite and obtain a lattice thermal conductivity $\kappa_l = 891$ W/mK at 300 K (with respect to a cross-sectional area of solid cylindrical model as mentioned before). Our result is consistent with previous studies of Zhang *et al.* [14] and Tan *et al.* [45] using the same definition of cross-sectional area. Both the quantum correction of MD temperature and the consideration of size effect ensure the correctness of our simulation results.

Summarizing all the calculated electron and phonon transport coefficients, we can now evaluate the ZT value. Fig. 2(d) plots the room temperature ZT value of (10, 0) tube as a function of carrier concentration. We see that the optimized ZT value is 0.15 for p-type system and 0.06 for n-type at almost identical concentration of $2.0\times10^{20}$ cm$^{-3}$. Although our calculated ZT values are higher than those measured experimentally [11, 12], they are still reasonable since we are dealing with a freestanding CNT with particular diameter, while the samples in the experiments are usually mixed with various types of CNTs which may influence their transport properties and thus the thermoelectric performance.

Up to now, we are dealing with room temperature. At elevated temperature, the lattice thermal conductivity can be reduced and thus a higher ZT value is expected. Figure 3 plots the lattice thermal conductivity as a function of temperature ranges from 300 K to 1000 K. It is found that the thermal conductivity decays exponentially with temperature. At 1000 K, the lattice thermal conductivity is calculated to be 390 W/mK, which is reduced by 56% compared with the room temperature value, and thus the thermoelectric performance could be greatly improved. The temperature dependence of ZT value is shown in Figure 4. We see that in the temperature ranges from 300 K to 1000 K, the ZT value keeps increasing for both p-type and n-type carriers, and reach the maximum of 0.77 and 0.35, respectively. Table II summarizes all the calculated ZT values and the corresponding transport coefficients. With increasing temperature, we see there is only a slight decrease of the power factor for both p-type and n-type carriers, while the electronic thermal conductivity keeps increasing. However, as the lattice thermal conductivity is much larger than the electronic part, the enhancement of ZT value is mainly attributed to the reduced lattice thermal conductivity with increasing temperature. To significantly improve the thermoelectric performance of CNT, the major effort should be thus directed to effectively reducing the lattice thermal conductivity. One of the good choices is the isotope substitution, which reduces the thermal conductivity of (10, 0) tube by 29% when half $^{12}$C atoms are substituted by $^{13}$C atoms [45]. Another effective strategy is surface design. It is found that chemisorptions of hydrogen on the outer surface of (10, 0) tube with a concentration of $C_{40}H_2$ could lower the lattice thermal conductivity by 48% [45]. If these two effects are combined, we find the lattice thermal conductivity of (10, 0) tube is decreased by 63%. Consequently, the *ZT* value can be significantly enhanced to 1.7 at 1000 K, which is one order of magnitude larger than the room temperature value, and is also comparable to those of the best thermoelectric material.

**4. Summary**

In summary, we demonstrate by multi-scale calculations that the (10, 0) CNT could become a high-performance thermoelectric material, even with an "unacceptable" high thermal conductivity. At room temperature, the predicted *ZT* value is very small, which is consistent with earlier studies. Nevertheless, the significantly reduced lattice thermal conductivity together with almost un-affected larger power factor lead to improved *ZT* value at elevated temperature. More importantly, the *ZT* value can be significantly enhanced to 1.7 at 1000 K by means of isotope substitution and hydrogen adsorption. The favorable thermoelectric performance of (10, 0) tube is guaranteed by its super electronic transport properties and an effective reduction of very high thermal conductivity. Our work emphasizes that major efforts should be directed towards the previously ignored mid- and high-temperature region, so that high thermoelectric performance could be realized in CNT containing earth-abundant and environmental-friendly elements.

**Acknowledgments**

We thank financial support from the National Natural Science Foundation (Grant No. 51172167 and J1210061) and the "973 Program" of China (Grant No. 2013CB632502)

**Table I** The relaxation time of (10, 0) SWCNT at 300 K. The corresponding carriers type, density-of-states effective mass $m^*$, elastic constant $C_{1D}$, and deformation potential constant $E_{1D}$ are also indicated.

| Carrier type | $C_{1D}$ (eV/ Å) | $m^*$ ($m_0$) | $E_{1D}$ (eV) | $\tau$ (s) |
|---|---|---|---|---|
| n | 521.24 | 0.282 | 8.203 | $6.58 \times 10^{-14}$ |
| p | 521.24 | −0.278 | −5.460 | $1.50 \times 10^{-13}$ |

**Table II** Calculated *ZT* values and corresponding transport coefficients at optimized carrier concentration for (10, 0) SWCNT operating at different temperatures.

| T (K) | Carrier type | n ($10^{20}$cm$^{-3}$) | $\tau$ ($10^{-14}$s) | L ($10^{-8}$WΩ/K$^2$) | S (μV/K) | σ ($10^7$S/m) | $S^2\sigma$ (W/mK$^2$) | $\kappa_e$ (W/mK) | $\kappa_l$ (W/mK) | ZT |
|---|---|---|---|---|---|---|---|---|---|---|
| 300 | p | 2.0 | 14.95 | 1.11 | 196 | 1.19 | 0.46 | 39.8 | 891 | 0.15 |
|  | n | 2.0 | 6.58 | 2.36 | −196 | 0.52 | 0.20 | 36.5 |  | 0.06 |
| 400 | p | 2.2 | 12.95 | 1.11 | 201 | 1.09 | 0.44 | 48.7 | 701 | 0.24 |
|  | n | 2.2 | 5.70 | 2.36 | −200 | 0.48 | 0.19 | 45.3 |  | 0.10 |
| 500 | p | 2.3 | 11.58 | 1.11 | 204 | 1.02 | 0.43 | 56.7 | 614 | 0.32 |
|  | n | 2.3 | 5.09 | 2.36 | −204 | 0.45 | 0.19 | 52.7 |  | 0.14 |
| 600 | p | 2.4 | 10.57 | 1.11 | 209 | 0.95 | 0.41 | 63.2 | 555 | 0.40 |
|  | n | 2.5 | 4.65 | 2.36 | −207 | 0.42 | 0.18 | 59.2 |  | 0.18 |
| 700 | p | 2.5 | 9.79 | 1.11 | 212 | 0.89 | 0.40 | 69.1 | 499 | 0.49 |
|  | n | 2.5 | 4.31 | 2.30 | −211 | 0.39 | 0.17 | 63.0 |  | 0.22 |
| 800 | p | 2.6 | 9.16 | 1.11 | 215 | 0.84 | 0.39 | 74.9 | 495 | 0.54 |
|  | n | 2.7 | 4.03 | 2.24 | −213 | 0.37 | 0.17 | 67.0 |  | 0.24 |
| 900 | p | 2.6 | 8.63 | 1.12 | 218 | 0.79 | 0.38 | 79.6 | 466 | 0.62 |
|  | n | 2.7 | 3.80 | 2.18 | −216 | 0.35 | 0.16 | 69.5 |  | 0.28 |
| 1000 | p | 2.6 | 8.19 | 1.12 | 223 | 0.73 | 0.36 | 81.8 | 390 | 0.77 |
|  | n | 2.7 | 3.60 | 2.11 | −221 | 0.33 | 0.16 | 69.2 |  | 0.35 |

.

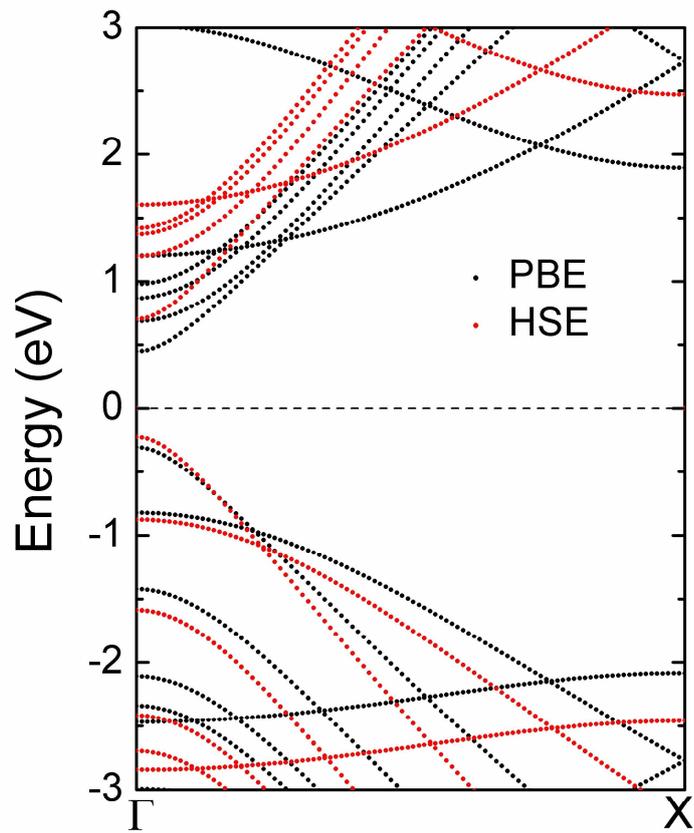

**Figure 1** The energy band structure of (10, 0) SWCNT. The black and red lines correspond to the calculations with PBE and HSE functional, respectively. The Fermi level is at 0 eV.

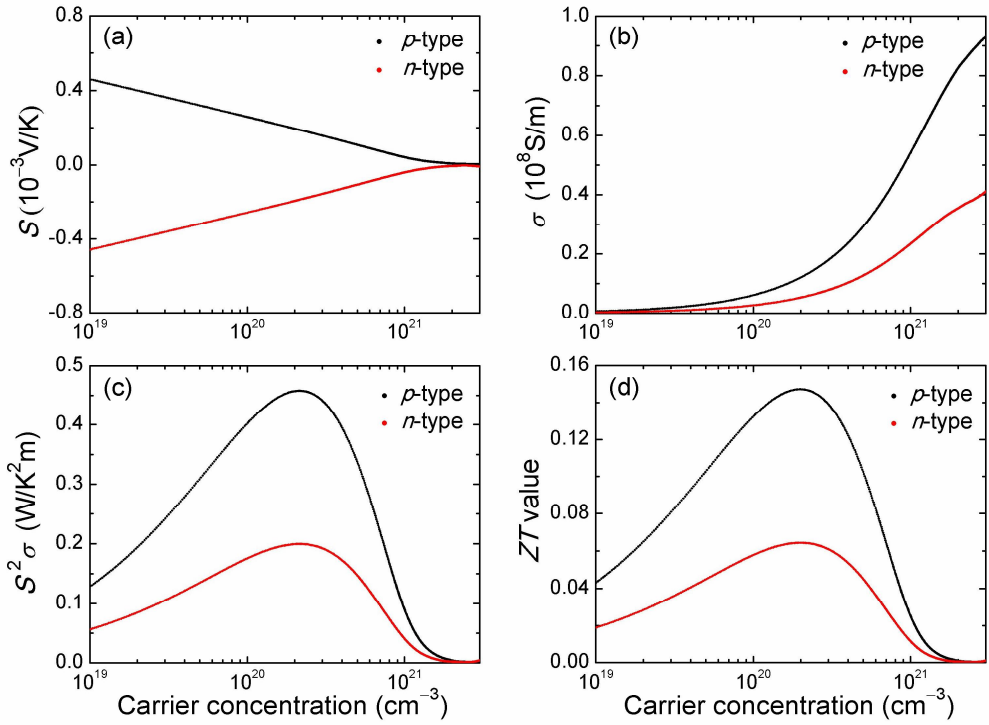

**Figure 2** The calculated electronic transport coefficients and *ZT* value as a function of carrier concentration for the (10, 0) SWCNT at 300 K: (a) the Seebeck coefficient, (b) the electrical conductivity, (c) the power factor, and (d) the *ZT* value.

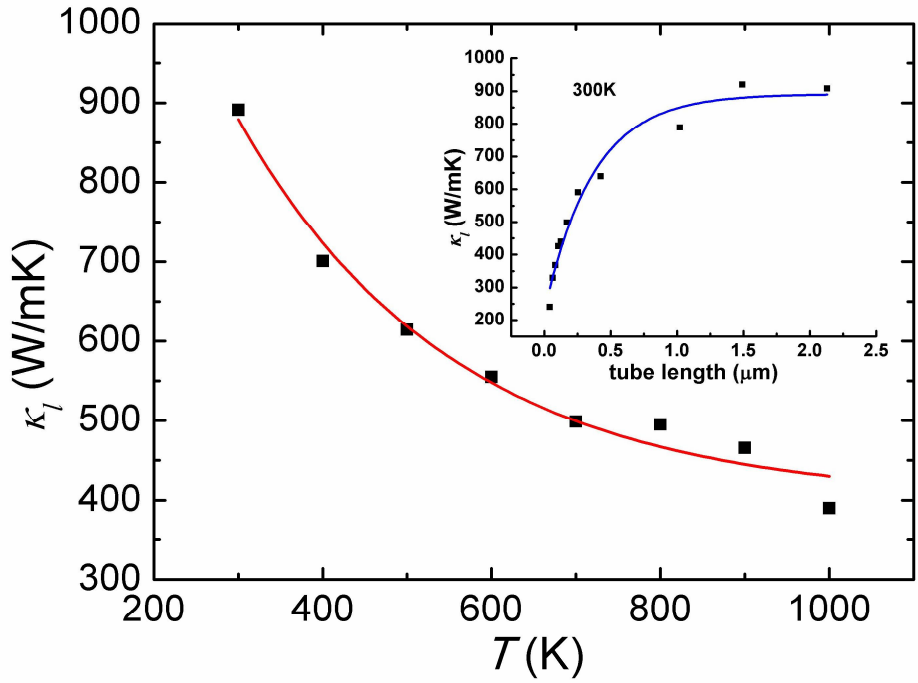

**Figure 3** The calculated lattice thermal conductivity of the (10, 0) SWCNT as a function of temperature. The red line represents an exponential fitting of the calculated results. The inset plots the lattice thermal conductivity as a function of tube length at 300 K, where the blue line indicates an exponential fitting.

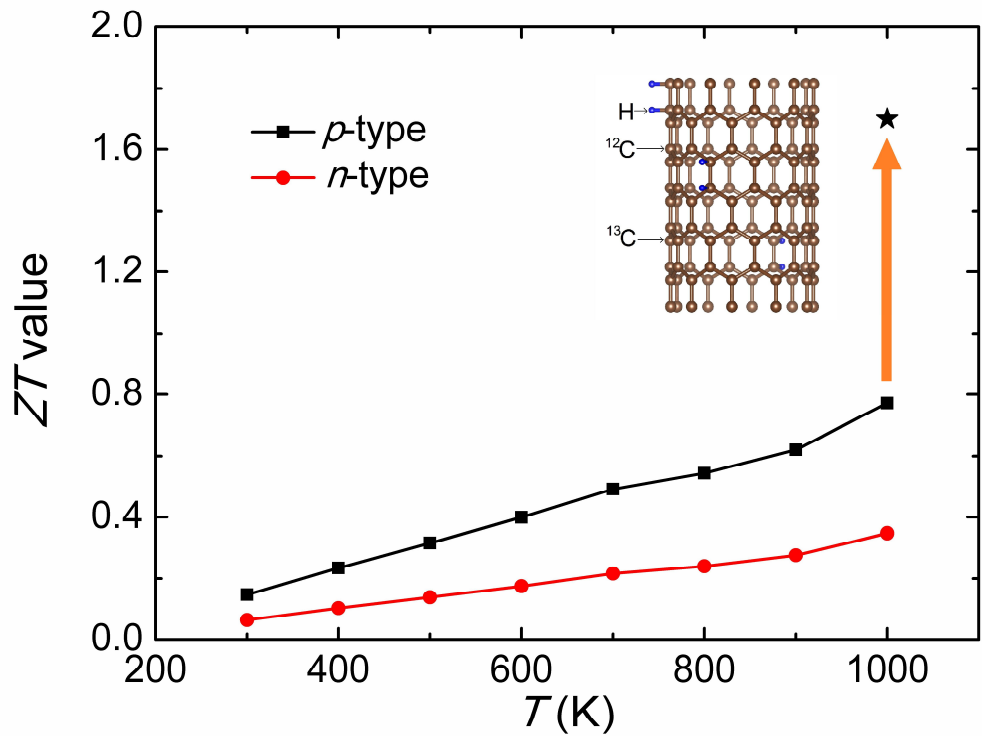

**Figure 4** The temperature dependence of *ZT* values for the (10, 0) SWCNT. The black and red lines correspond to *p*- and *n*-type carriers, respectively. The arrow indicates that the *ZT* value at 1000 K can be significantly enhanced by isotope substitution and chemisorptions of hydrogen, as shown in the inset model.